\documentclass[a4paper]{article}
\usepackage[cp866]{inputenc}
\usepackage{graphicx}
\date{}
\author{S.\,V.\,\,Sazonov\thanks{e-mail address: barab@newmail.ru}, 
\,and N.\,V.\,\,Ustinov$\mbox{}^{a,}$\thanks{e-mail address: n\_ustinov@mail.ru}
\medskip\\
Theoretical Physics Department, Kaliningrad State University,\\\medskip
A.\,Nevskogo str.\,\,14, Kaliningrad, 236041, Russia\\
$\mbox{}^a$Quantum Field Theory Department, Tomsk State University,\\
36 Lenin Avenue, Tomsk, 634050, Russia
\vspace{-4ex}}
\title{Optical transparency modes in anisotropic media}
\date{ }
\begin{document}
\maketitle
\begin{abstract}
\noindent
The modes of nonlinear propagation of the two-component electromagnetic pulses
through optically uniaxial media containing resonant particles are studied.
The features of their manifestation in the ''dense'' media and in the media
with expressed positive and negative birefringences are discussed.
It is shown that exponentially and rationally decreasing solutions of the
system of material and wave equations allow us also to describe the 
propagation of the self-induced transparency pulses in isotropic media in the 
case, when the direct electric dipole-dipole interaction between the resonant 
particles is taken into account.
\end{abstract}


\section{INTRODUCTION}

The development of the technologies of producing the low-dimensional quantum
structures (wells, wires), growing the semiconductor crystals strengthens an 
interest to theoretical investigation of the nonlinear coherent phenomena in 
the anisotropic media$\mbox{}^{1-8}$.
The quantum states of the particles being contained in such the media possess 
no certain parity, owing to what the diagonal elements of the matrix of the 
dipole moment operator and their differences, which are called the permanent 
dipole moments of the transitions, are distinct from zero.

The propagation in optically uniaxial media of the two-component laser pulses
consisting of ordinary and extraordinary components was studied
in$\mbox{}^{7,8}$.
It was suggested$\mbox{}^8$ that the linear velocities of both the components
are equal, and the concentration of the resonant particles is small 
sufficiently for the approximation of the unidirectional propagation to be 
valid.
Various modes of the optical transparency were classified with regard to the
pulse propagation velocity and the degree of the resonant particles
excitation.
The mode of self-induced transparency (SIT) is characterized by strong
excitation of the medium and substantial deceleration in the velocity of the
pulse propagation relative to linear one.
The self-induced supertransparency (SIST) mode differs from SIT in that the
decrease of velocity of the pulses is small, but the medium is strongly 
excited as well.
In this mode, the carrier frequency of ordinary component of the pulses, which 
cause full inversion of the quantum level population, is lower than the 
resonance frequency.
The width of such the pulses is determined by the amount of the ordinary 
component detuning.
(The full inversion of initial state of the resonant particles in the 
isotropic case happens under condition of the exact resonance only.)
The SIST pulses have larger amplitudes and smaller duration as compared to
the SIT ones, and local frequency of their ordinary component is strongly 
modulated.
There exist also the modes in that the trapping of the quantum level 
population takes place.
The pulses propagating in the extraordinary transparency (EOT) mode are
characterized by small detuning of the ordinary component from the resonance 
and by dominant role of the extraordinary component.
Their group velocity substantially changes and may become comparable to one of 
the pulses causing strong excitation.
In the modes of positive nonresonant transparency (PNT) and negative
nonresonant transparency (NNT), the pulse velocity changes insignificantly,
and the absolute value of detuning is large.
The extraordinary component of the pulse dominates in the former mode, whereas
the ordinary-to-extraordinary amplitude ratio is arbitrary in the latter.
The most substantial difference between these modes concerns the behavior of
effective detuning.
In the NNT mode, it remains virtually constant.
If a pulse propagates in the PNT mode, then the effective detuning of the
frequency of the ordinary component changes sign due to an influence of the 
extraordinary one.
Since the local frequency passes through the resonance, a slightly stronger 
excitation of the medium occurs here, and the pulses are sharply peaked, as in 
the SIST mode.

In present report, we consider the propagation of the two-component
electromagnetic pulses through the optically uniaxial medium containing 
two-level particles. 
Unlike the case studied in$\mbox{}^8$, the approximation of unidirectional 
propagation is not applied.
Also, linear velocities of ordinary and extraordinary components are not
assumed to be equal.
In Section 2, we write down the system of material and wave equations for
resonant propagation of the electromagnetic pulses in a direction 
perpendicular to the optical axis of the anisotropic medium.
Exponentially and rationally decreasing solutions of this system are given here 
as well.
Using these solutions, we establish in Sections 3 and 4 the distinctive 
features of the manifestation of the transparency modes under discussion in 
''dense'' media and in the media with expressed positive and negative 
birefringences.
The pulse solutions studied in the previous sections are modified in Section 5
on the case of isotropic, optically dense media, in which the direct electric
dipole-dipole interaction of the resonant particles should be taken into 
account.

\section{BASIC EQUATIONS AND THEIR SOLUTIONS}

Consider a birefringent medium containing the resonant, two-level particles. 
Suppose that the plane, two-component pulse consisting of high-frequency 
ordinary component $E_o$ and video-pulse extraordinary one $E_e$, propagates 
through the medium in a direction perpendicular to its optical axes.
The system of material and wave equations describing the dynamics of the
quantum systems and the evolution of the components of the electric field 
in the slowly varying envelope approximation has the next form$\mbox{}^7$:
\begin{equation}
\frac{\partial W}{\partial t}=\frac{i}{2}\left(\Omega^*_oR-\Omega_oR^*\right),
\label{1}
\end{equation}
\begin{equation}
\frac{\partial R}{\partial t}=i\left(\Delta+\Omega_e\right)R+i\Omega_oW,
\label{2}
\end{equation}
\begin{equation}
\frac{\partial\Omega_o}{\partial x}+\frac{n_o}{c}
\frac{\partial\Omega_o}{\partial t}=-i\beta_oR,
\label{3}
\end{equation}
\begin{equation}
\frac{\partial^2\Omega_e}{\partial x^2}-\frac{n_e^2}{c^2}
\frac{\partial^2\Omega_e}{\partial t^2}=-2\beta_e\frac{n_e}{c}
\frac{\partial^2W}{\partial t^2}.
\label{4}
\end{equation}
Here the inversion population $W$ ($-1/2\le W\le1/2$) and coherency $R$ are 
expressed through the elements of the density matrix of the resonant 
particles; $\Delta=\omega_0-\omega$ is detuning of the carrier frequency 
$\omega$ of the ordinary component from resonant frequency $\omega_0$ of the 
quantum transition ($|\Delta|\ll\omega_0$); $\Omega_o=2d{\cal E}_o/\hbar$; 
${\cal E}_o$ is an envelope of the pulse ordinary component 
($E_o={\cal E}_o\exp(i\omega(t-n_ox/c))+{\rm c.c.}$); $\Omega_e=DE_e/\hbar$; 
$d$ and $D$ are the dipole moment and the permanent dipole moment of the 
transition, respectively; $\hbar$ is the Plank's constant; 
$\beta_o=4\pi Nd^2\omega/(\hbar cn_o)$; $\beta_e=2\pi ND^2/(\hbar cn_e)$;
$N$ is the concentration of the particles; $c$ is the velocity of the light in
vacuum; $n_o$ and $n_e$ are ordinary and extraordinary refractive indices; 
$x$ and $t$ are the physical distance and time.

The nonlinear equations arising from system (\ref{1}--\ref{4}) under condition
$n_e=n_o$ in the approximation of unidirectional propagation and their pulse 
solutions were investigated in work$\mbox{}^8$.
The classification of the modes of nonlinear propagation of the two-component 
electromagnetic pulses, which are distinguished by the behavior of the quantum 
particles and by the characteristics of the pulses, was given.
It turns out that the pulse solutions of the system of material and wave
equations studied there can be modified properly for the case considered.
Indeed, the following expressions are obtained
\begin{equation}
\Omega_o=\sqrt{M}\exp(i\Phi),
\label{5}
\end{equation}
\begin{equation}
\Omega_e=-\frac{\tilde D}{4d^2\omega}\,M,
\label{5a}
\end{equation}
\begin{equation}
W=\left(1-\frac{\tau_p^2}{2(1+\alpha^2)}\,M\right)W_{\infty},
\label{6}
\end{equation}
where 
\begin{equation}
M=\frac{8g}{\tau_p^2\left(g-\alpha+\mbox{sgn}(g)\sqrt{1+(g-\alpha)^2}
\cosh(2\zeta)\right)^{\mathstrut}};
\label{7}
\end{equation}
\begin{equation}
\Phi=W_{\infty}\frac{\beta_o\alpha\tau_p}{1+\alpha^2}\,x-\mbox{arctan}
\frac{\tanh\zeta}{s}+{\rm const};
\label{8}
\end{equation}
\begin{equation}
\tilde D=\frac{A_e}{\displaystyle\left(1+\frac{n_o-n_e}{n_o+n_e}
\frac{1+\alpha^2}{A\tau_p^2}\right)\left(1+\frac{A\tau_p^2}{1+\alpha^2}
\right)^{\mathstrut}};
\label{9}
\end{equation}
$$
\alpha=\Delta\tau_p;\quad g=2\omega_0\tau_p\frac{d^2}{\tilde D};\quad
\zeta=\frac{t}{\tau_p}-\frac{x}{v_g\tau_p};
$$
$$
A_e=\frac{2n_oD^2}{n_o+n_e};\quad
A=-\frac{c\beta_oW_{\infty}}{n_o+n_e};
$$
$$
s=g-\alpha+\mbox{sgn}(g)\sqrt{1+(g-\alpha)^2};\quad
v_g=\frac{c}{\displaystyle n_o+\frac{n_o+n_e}{1+\alpha^2}A\tau_p^2};
$$
$\tau_p$ is positive parameter; $W_\infty$ is an initial population of the 
quantum levels.
We will assume in the subsequent sections that the initial state of the 
quantum particles is stable: $-1/2\le W_{\infty}<0$.

It is seen that the ordinary component phase modulation
$\displaystyle\frac{\partial \Phi}{\partial t}$ is equal to
$\displaystyle\frac{\Omega_e}{4}$.
Defining the pulse duration $T_p$ as the double deviation from the zero point
of $t-x/v_g$, at which $|\Omega_o|$ is half its maximum value, we obtain from
formula (\ref{5}):
$$
T_p=\tau_p\,{\rm arccosh}\left(4+3\,\mbox{sgn}(g)\frac{g-\alpha}{
\sqrt{1+(g-\alpha)^2}}\right).
$$

Since parameter $\tilde D$ depends not only on the characteristics of the
medium, as it was$\mbox{}^8$, but also from parameters $\Delta$ and $\tau_p$
of the pulse, the anisotropy caused by the permanent dipole moment becomes
effective.
Thereof, the domains of the pulse parameters values, in which exist the 
optical transparency modes described in$\mbox{}^8$, will change.
It is remarkable that the condition of strong excitation of the medium is
written in the same manner:
\begin{equation}
g=\frac12\left(\alpha+\frac{1}{\alpha}\right).
\label{10}
\end{equation}
The pulses, whose parameters satisfy this condition, will cause the full
inversion of initial state of the quantum particles.
The parameter $|\tilde D|$, which does not vanish only for a medium having a
permanent dipole moment, will be called the effective anisotropy in the sequel.

Taking the limit $\tau_p\to\infty$ in formulas (\ref{5}--\ref{9}), we obtain
rationally decreasing solution of system (\ref{1}--\ref{4}):
\begin{equation}
\Omega_o=\frac{8i\omega_0d^2\kappa}{\tilde D(1+i\kappa^2b)}\exp(iax),
\label{11}
\end{equation}
\begin{equation}
W=\left(1-\frac{8\kappa^2}{(1+\kappa^2)^2(1+\kappa^4b^2)}\right)W_{\infty},
\label{11a}
\end{equation}
where
$$
a=\frac{\beta_oW_{\infty}}{\Delta};\quad
b=4\omega_0\frac{d^2}{\tilde D}\left(t-\frac{x}{v_g}\right);
$$
$$
\kappa=\sqrt{\frac{\Delta\hat D}{2\omega_0d^2}-1};\quad
v_g=\left(\frac{n_o}{c}-\frac{\beta_oW_{\infty}}{\Delta^2}\right)^{-1};
$$
$$
\tilde D=\frac{A_e}{\displaystyle\left(1+\frac{n_o-n_e}{n_o+n_e}
\frac{\Delta^2}{A}\right)^{\mathstrut}\left(1+\frac{A}{\Delta^2}\right)}.
$$
This solution, as the algebraic solution of the derivative nonlinear
Schr\"odinger equation$\mbox{}^9$ (DNSE), has one parameter (without taking
account of the shifts on variables $x$ and $t$).
This is not surprising, because the system of the equations considered
in$\mbox{}^8$ and DNSE belong to the same hierarchy of the nonlinear equations
integrable in the frameworks of the inverse scattering transformation
method$\mbox{}^{10-12}$.
Also, rationally decreasing solution having no arbitrary  parameters was found
for the case of the one-component pulses propagating through the anisotropic
media$\mbox{}^6$.

Parameter $\kappa$ in formulas (\ref{11}) and (\ref{11a}) is supposed to be
real.
This imposes a constraint on detuning $\Delta$ of the rationally decreasing
pulses.
It is seen from equation (\ref{11a}) that the full inversion of the medium
takes place if detuning of the pulse satisfy condition $\kappa=1$.

\section{THE CASE OF ''DENSE'' MEDIA}

In this section, we discuss the case of ''dense'' media, when the 
concentration of the resonant particles is large enough so that condition 
$A\tau_p^2\ll1$ is not satisfied.
We use quotes to stress its difference from the case of optically dense media,
in which the effects of the local field are considered.
Since the refractive indices of typical anisotropic medium are such that
inequality $|n_e-n_o|\ll1$ holds, we can assume that the first multiplicand in
the denominator in the right-hand side of (\ref{9}) is close to unity.
Then, formula (\ref{9}) is rewritten as
\begin{equation}
\tilde D=\frac{A_e}{\displaystyle1+\frac{A\tau_p^2}{1+\alpha^2}}.
\label{D_dm}
\end{equation}

It can be shown that the transparency modes described in$\mbox{}^8$ exist for
the ''dense'' media as well.
If $n_e<n_o$, then $\tilde D$ can exceed $D^2$ in the general case.
One can see that this takes place if
$$
\frac{1}{\tau_p^2}+\Delta^2>\frac{n_o+n_e}{n_o-n_e}\,A.
$$
However, expression (\ref{9}) cannot be replaced by (\ref{D_dm}) under this
condition.
Therefore, the effective anisotropy cannot exceed $D^2$ in the ''dense''
medium case, and the modes of strong excitation (with $g>1$) must be more
expressed.

Using (\ref{10}) and (\ref{D_dm}), we find that parameters $\tau_p$ and 
$\Delta$ of the pulse causing the full inversion of the medium are related as 
given:
\begin{equation}
\frac{1}{\tau_p}=\sqrt{F_{1,2}}\,,
\label{tau_d}
\end{equation}
where
\begin{equation}
F_{1,2}=-\Delta^2+\left(\Delta\pm\sqrt{\Delta\Bigl(\Delta+AA_e/(\omega_0d^2)
\Bigr)}\right)\frac{2\omega_0d^2}{A_e}.
\label{F_12}
\end{equation}
It is obvious that
\begin{equation}
F_1F_2=\Bigl(A_e\Delta^3/(4d^2)-\omega_0\Delta^2-A\omega_0\Bigr)
\frac{4d^2\Delta}{A_e}.
\label{*}
\end{equation}

It follows from (\ref{tau_d}) that detuning of the pulses, whose propagation 
is accompanied  by the largest change of the level population, can only be 
such that at least one of the quantities $F_1$ and $F_2$ is positive.
It is easy to see that this is possible only if $\Delta>0$.
Since the coefficients at $\Delta^3$ and $\Delta$ in the right-hand side of
(\ref{*}) are positive and the coefficient at $\Delta^2$ is zero, the product
$F_1F_2$ changes sign only once at $\Delta>0$.
Therefore, $F_2$ is negative in this domain, and $F_1$ is positive if
$0<\Delta<\Delta_d$, where $\Delta_d$ is the positive root of equation
$F_1F_2=0$.
Thus, for a given detuning satisfying condition $0<\Delta<\Delta_d$, only
the pulses with a certain unique duration can strongly excite the medium.
A similar result was obtained in the case, when the unidirectional propagation 
approximation was used, and the refractive indices were assumed to be
equal$\mbox{}^8$.
The carrier frequency of the ordinary component of a pulse that strongly
excites the medium was also lower than the resonance frequency.
Note that parameter $\tau_p$ cannot be smaller than a certain minimum for such
the pulses.
Furthermore, in accordance with the classification$\mbox{}^8$, both
modes of strong excitation can be identified in this case: the SIT
($\alpha\to0$) and SIST ($\alpha\to\infty$) modes correspond to $\Delta\to0$
and $\Delta\to\Delta_d$ respectively (see Fig.~1).

\begin{figure}[ht]
\centering
\vskip-2.1cm
\includegraphics[angle=-90,width=0.9in]{A_00.eps}
\includegraphics[angle=-90,width=4.0in]{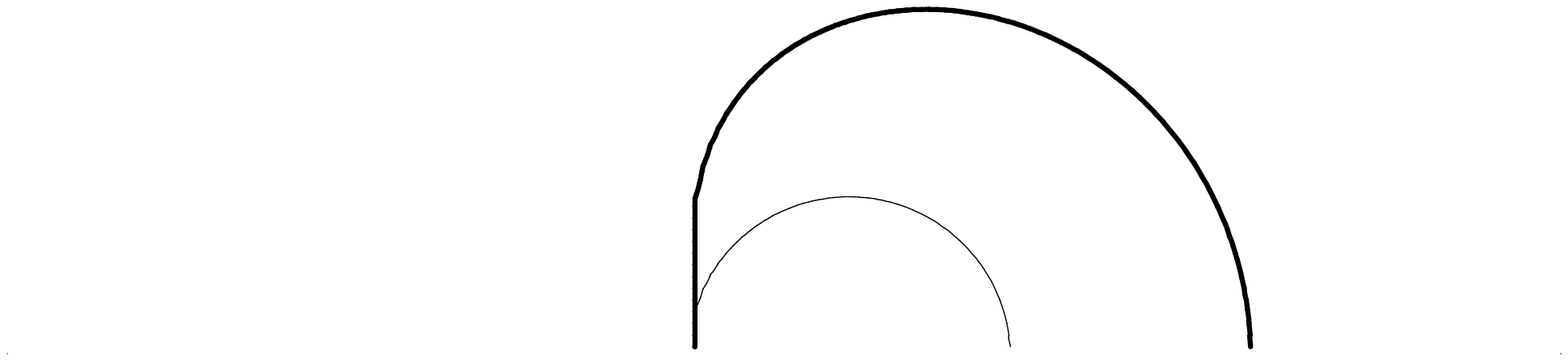}
\begin{picture}(80,40)
\put(-168,-130){\small I}
\put(-100,-130){\small II}
\put(-176.2,-78){\small III}
\put(-246,-120){\small IV}
\put(-36,-120){\small V}
\put(-260,-133.5){\vector(1,0){240}}
\put(-160.7,-133.4){\vector(0,1){100}}
\put(-27,-146){$\Delta$}
\put(-163.2,-146){0}
\put(-92.5,-146){$\Delta_d$}
\put(-124,-146){$\Delta_m$}
\put(-182,-44){$\tau_p^{-1}$}
\end{picture}
\vskip-2.4cm
\centerline{
\parbox{14.0cm}{\small\bf Fig. 1. \rm Curve of the strong excitation for
''dense'' medium. Domains of the pulse parameters are given in the case
$\Delta_d\ll\omega_0$: (I) SIT; (II) SIST; (III) EOT; (IV) NNT; (V) PNT.
Thin curve: $A=0$.}}
\vskip-0.1cm
\end{figure}

Let us note that the formulas, which appear in the frameworks of the
unidirectional propagation approximation under condition $n_e=n_o$, are
obtained from equations (\ref{tau_d}) and (\ref{F_12}) if $A=0$.
The curve of the strong excitation for this case is depicted on Fig.~1 by thin
line.
The corresponding value of the maximum detuning is
$\Delta_m=4\omega_0d^2/D^2$.
One can show that the interval of admissible detuning is wider for the
''dense'' medium ($\Delta_d>\Delta_m$).
This can be explained as follows.
The effective anisotropy is more expressed in a low-density resonant medium,
and a pulse can be resonant with the medium on average of its duration and
thus cause strong excitation of the medium within a narrower interval of
detuning.
Note also that the medium effectively becomes more isotropic with increasing
$A$, and also the slope of the curve described by (\ref{tau_d}) increases with
decreasing $\Delta$.
This is a consequence of the fact that the largest change of the quantum-level
population in an isotropic medium can occur under the exact resonance 
condition only.

It follows from the consideration carried out above that the decrease in the
effective anisotropy in the ''dense'' medium does reinforce characteristic
properties of pulses in the SIT and SIST modes.
Conversely, the NNT, PNT, and EOT modes (with $g\ll1$) become less expressed.
The difference between the refractive indices of anisotropic ''dense'' medium 
is not essential.
The domains of the pulse parameters values in different modes of optical
transparency are marked on Fig.~1.

The expression for parameter $\tilde D$ of the rationally decreasing pulse
(\ref{11}), (\ref{11a}) is obtained from formula (\ref{D_dm}) in limit
$\tau_p\to\infty$.
The interval of possible values of detuning of this pulse is determined by the
condition of reality of parameter $\kappa$.
It is easy to show that detuning of rationally decreasing pulse should be
positive and can be smaller than $\Delta_d$.

\section{MEDIA WITH EXPRESSED BIREFRINGENCES}

Suppose that the concentration of the resonant particles is so small that
inequality $A\tau_p^2\ll1$ is valid.
Then we have from equation (\ref{9}):
\begin{equation}
\tilde D=\frac{A_e}{\displaystyle1+\frac{n_o-n_e}{n_o+n_e}
\frac{1+\alpha^2}{A\tau_p^2}}.
\label{9a}
\end{equation}
In contrast to the case of ''dense'' medium considered in the previous
section, the difference of refractive indices plays significant role here.

Let us assume at first that the medium is positively birefringent ($n_e>n_o$).
In this case, $\tilde D$ is negative if the pulse parameters $\tau_p$ and
$\Delta$ satisfy condition
$$
\frac{1}{\tau_p^2}+\Delta^2>\frac{n_e+n_o}{n_e-n_o}\,A.
$$
The polarity of the extraordinary component changes the sign here (see formula
(\ref{5a})).
This leads to important consequences concerning the modes of the strong
excitation of the medium.

As $\tau_p\to\tilde\tau_p$ and $\Delta\to\tilde\Delta$, where $\tilde\tau_p$
and $\tilde\Delta$ are connected by relation
$$
\frac{1}{{\tilde\tau_p}^2}+{\tilde\Delta}^2=\frac{n_e+n_o}{n_e-n_o}\,A,
$$
the effective anisotropy increases indefinitely ($|\tilde D|\to\infty$), and
the group velocity of the pulse tends to the extraordinary-wave velocity
($v_g\to c/n_e$).
A pulse with $\tau_p$ and $\Delta$ close to $\tilde\tau_p$ and $\tilde\Delta$
propagates in the EOT mode (see Fig.~2), as its extraordinary component is
stronger than the ordinary one.
Since the effective anisotropy is larger in the parameter domain in question,
the EOT regime is more expressed, especially when $\Delta$ is small.

\begin{figure}[ht]
\centering
\vskip-1.0cm
\includegraphics[angle=-90,width=4.0in]{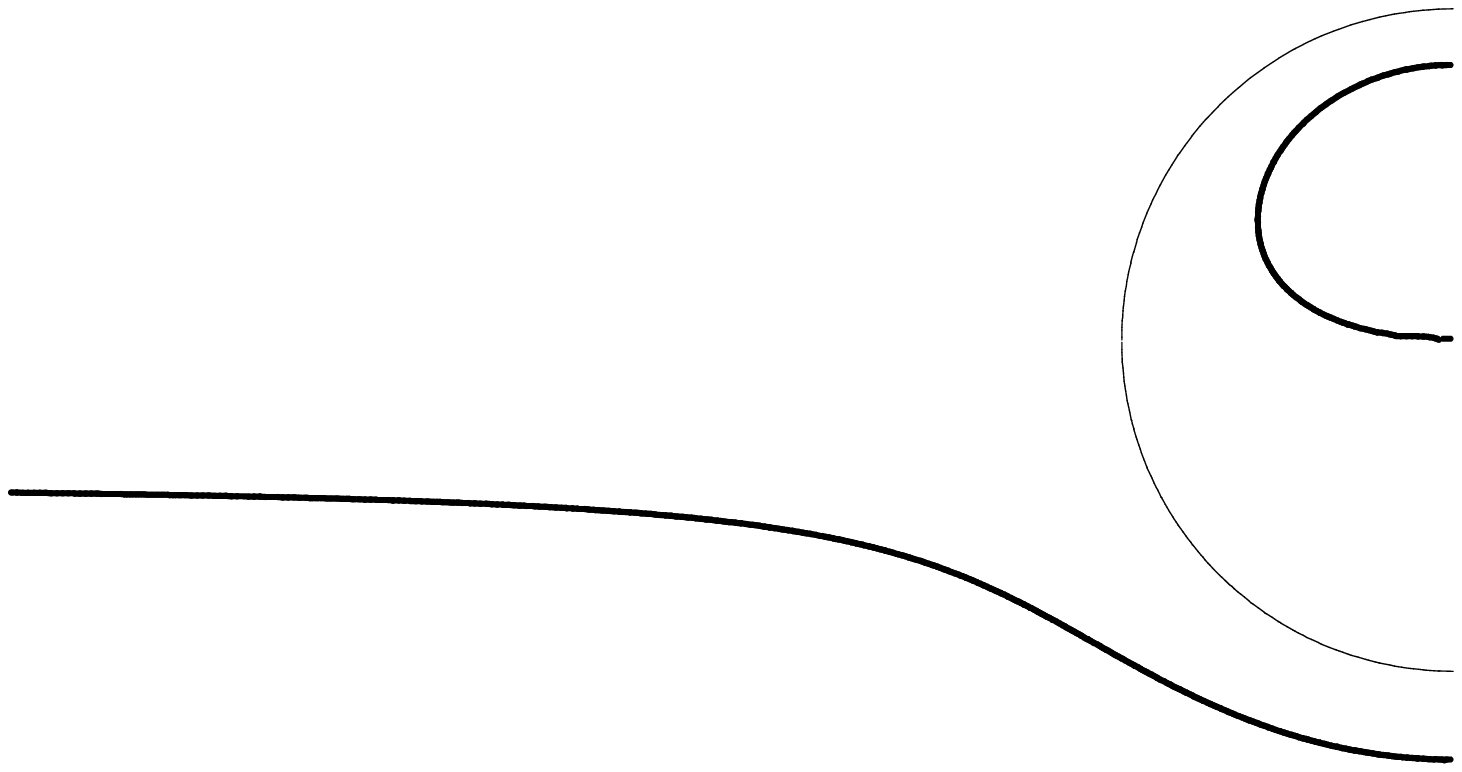}
\begin{picture}(0,0)
\put(-149,-185){\small I}
\put(-168,-44){\small I}
\put(-126,-185){\small II}
\put(-198,-180){\small II}
\put(-157,-164){\small III}
\put(-126,-154){\small III}
\put(-266,-176){\small V}
\put(-49,-176){\small IV}
\put(-272.5,-187.5){\vector(1,0){240}}
\put(-144.1,-187.5){\vector(0,1){154}}
\put(-38,-200){$\Delta$}
\put(-146.5,-200){0}
\put(-119.5,-200){$\Delta_1$}
\put(-164.5,-200){$\Delta_0$}
\put(-193,-200){$\Delta_2$}
\put(-138,-42){$\tau_p^{-1}$}
\put(-160,-189){\line(0,1){4}}
\put(-130,-189){\line(0,1){4}}
\put(-134.5,-200){$\tilde\Delta_1$}
\put(-210,-189){\line(0,1){4}}
\put(-214.5,-200){$\tilde\Delta_2$}
\put(-108.5,-189){\line(0,1){4}}
\put(-106,-184){$\Delta_+$}
\put(-178.5,-189){\line(0,1){4}}
\put(-176,-184){$\Delta_-$}
\end{picture}
\vskip-0.5cm
\centerline{
\parbox{14.0cm}{\small\bf Fig. 2. \rm Curve of the strong excitation for
positively birefringent medium. Parameter domains are shown for
$|\Delta_2|\ll\omega_0$ and labelled as on Fig.~1. Thin curve: $v_g=c/n_e$.}}
\vskip-0.2cm
\end{figure}

If
\begin{equation}
\frac{1}{\tau_p^2}+\Delta^2>\frac{n_e+3n_o}{n_e-n_o}\,A,
\label{c}
\end{equation}
then the effective anisotropy cannot exceed $D^2$.
Therefore, the NNT and PNT modes, for which $g\ll1$ and $|\alpha|\gg1$, are
less expressed in this parameter domain.
Since $\tilde D<0$ here, the detuning corresponding to these modes changes
the sign; i.e., the former regime exists at $\alpha<0$; the latter, at
$\alpha>0$.

Consider the case of strong excitation of the medium with positive
birefringence.
It follows from (\ref{10}) and (\ref{9a}) that the largest change of the
population of the quantum levels occurs if the pulse parameters $\tau_p$ and
$\Delta$ are such that
\begin{equation}
\frac{1}{\tau_p}=\sqrt{F},
\label{tau_+-}
\end{equation}
where
\begin{equation}
F=\frac{\omega_0\Delta}{A_1+A_2\Delta}-\Delta^2;
\label{F_+-}
\end{equation}
$$
A_1=\frac{A_e}{4d^2};\quad
A_2=\frac{n_e-n_o}{n_e+n_o}\,\frac{\omega_0}{A}.
$$
Condition $F>0$ determines the range of admissible detuning.
Using (\ref{tau_+-}), we rewrite (\ref{9a}) as
\begin{equation}
\tilde D=4d^2\left(A_1+A_2\Delta\right).
\label{D_+-}
\end{equation}
We denote by
$$
\Delta_0=-\frac{A_1}{A_2},
$$
and
\begin{equation}
\Delta_{1,2}=\left(-A_1\pm\sqrt{A_1^2+4A_2\omega_0}\right)\frac{1}{2A_2}.
\label{Delta_+-}
\end{equation}
the root of equation $\tilde D=0$ and the nonzero roots of equation $F=0$.
Since $n_e>n_o$ in the case considered, it holds that
$\Delta_2<\Delta_0<\Delta_1$ (see Fig.~2).

One can see that the largest change in the quantum level population can be 
achieved for both positive and negative values of detuning 
($0<\Delta<\Delta_1$ and $\Delta_2<\Delta<\Delta_0$).
The existence of the latter interval of admissible detuning is due to the
possibility of change in the sign of $\tilde D$ for positively birefringent
media.
Indeed, if $\Delta_2<\Delta<\Delta_0$, then $\tilde D<0$ and therefore
$\Omega_e>0$.
This means that a negatively detuned pulse can strongly excite the medium,
being resonant with it on average of the pulse duration because of the dynamic
shift induced by its extraordinary component.

The transparency modes accompanied by the strong excitation of the medium in
the case under consideration are similar to those described in$\mbox{}^8$, but
they can exist in two parameter domains.
If $\Delta\to0$ or $\Delta\to\Delta_0$, then $\alpha\to0$, and the pulse
propagates in the SIT mode.
In the latter limit case, both components have larger amplitudes and,
therefore, higher velocities.
If $\Delta\to\Delta_1$ or $\Delta\to\Delta_2$, then $|\alpha|\to\infty$, and
the pulse propagates in the SIST mode, in which the pulse shape is sharper
and velocity is higher as compared to the case when $\Delta\to0$.
The corresponding transparency modes domains and the curve of strong
excitation are shown on Fig.~2.

If $n_e\to n_o$, then $\Delta_1\to\omega_0/A_1$, and both $\Delta_0$ and
$\Delta_2$ increase indefinitely.
The curve of strong excitation for this case is represented in Fig.~1 by a
thin line.

It should be noted that positively birefringent media, as well as isotropic
media, are strongly excited by pulses with arbitrary $\tau_p$.
Since the present analysis relies on condition $A\tau_p^2\ll1$, the effective
anisotropy exceeds $D^2$ in the parameter domain where inequality
(\ref{c}) is not valid.
Therefore, both SIT and SIST modes are less expressed in the case of positive
detuning, whereas the SIT mode corresponding to $\Delta\to\Delta_0$ must be
more expressed because the effective anisotropy tends to zero.

Taking limit $\tau_p\to\infty$ in formula (\ref{9a}), we obtain expression for
parameter $\tilde D$ of rationally decreasing pulse (\ref{11}), (\ref{11a}) in
the media with expressed birefringences.
In positively birefringent media, detuning of this pulse can be both positive:
$\tilde\Delta_1<\Delta<\Delta_+$, and negative:
$\tilde\Delta_2<\Delta<\Delta_-$.
Here we use notations
$$
\tilde\Delta_{1,2}=\left(-A_1\pm\sqrt{A_1^2+A_2\omega_0}\right)\frac{1}{A_2};
$$
$$
\Delta_{\pm}=\pm\sqrt{\frac{\omega_0}{A_2}}.
$$
Detunings $\Delta_1$ and $\Delta_2$ belong to the ranges of possible values of
detuning for such the pulses (see Fig.~2).

Next, we consider negatively birefringent media ($n_o>n_e$).
It is easy to show by using condition $A\tau_p^2\ll1$ that $\tilde D<D^2$;
i.e., the effective anisotropy is weaker for such media.
The reason is that the linear extraordinary-wave velocity is higher as
compared to the ordinary one and, therefore, is always higher than the pulse
velocity.
In this case, the role of the extraordinary component in the pulse formation
occurs to be less significant.

\begin{figure}[ht]
\centering
\vskip-1.0cm
\includegraphics[angle=-90,width=4.0in]{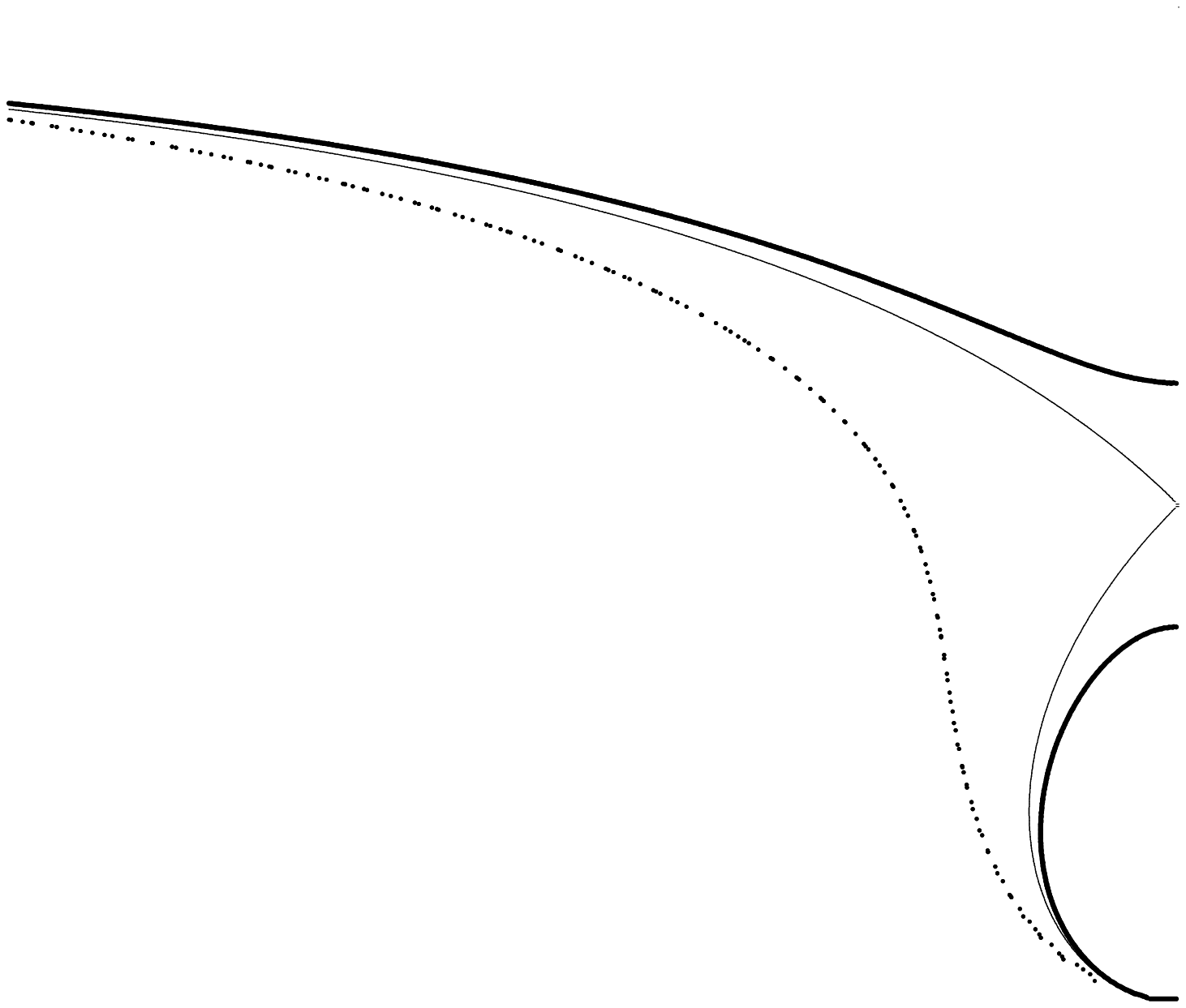}
\begin{picture}(0,0)
\put(-256,-187.5){\vector(1,0){224}}
\put(-213,-187.5){\vector(0,1){154}}
\put(-39,-200){$\Delta$}
\put(-215.5,-200){0}
\put(-169.5,-200){$\Delta_1$}
\put(-89,-200){$\Delta_0$}
\put(-137.5,-200){$\Delta_2$}
\put(-234.5,-42){$\tau_p^{-1}$}
\put(-84.5,-189){\line(0,1){4}}
\end{picture}
\vskip-0.5cm
\centerline{
\parbox{14.0cm}{\small\bf Fig. 3. \rm Curves of the strong excitation of
negatively birefringent medium for $-A_2\omega_0<A_1^2/4$,
$-A_2\omega_0=A_1^2/4$ (thin curve), and $-A_2\omega_0>A_1^2/4$
(dotted curve).}}
\vskip-0.2cm
\end{figure}

Since negative birefringence reduces the effective anisotropy, the modes
of transparency characterized by the strong excitation are more expressed.
Equations (\ref{tau_+-}--\ref{Delta_+-}) corresponding to these modes remain
valid.
Note that condition $A\tau_p^2\ll1$ imposes a constraint on parameter $\tau_p$
and, therefore, on detuning (see (\ref{tau_+-})).
Accordingly, the detuning in (\ref{D_+-}) must be such that parameter
$\tilde D$ is less than $D^2$ even though the first summand in the righthand
side is greater than $D^2$.

It is seen from formula (\ref{F_+-}), that $F<0$ for negative $\Delta$; i.e.,
the pulses that cause the largest change of the level population can be 
positively detuned only.
This agrees with the fact that $\tilde D$ does not change sign in the case
under study, and, therefore, the extraordinary component induces a dynamic red
shift.
However, as in the case of positive detuning, two intervals of admissible
detuning corresponding to strong excitation of the medium can exist.
This takes place if the medium is such that
$0<-A_2\omega_0<A_1^2/4$.
In this case, $\Delta_1$ and $\Delta_2$ are real and distinct, at that
$\Delta_1<\Delta_2<\Delta_0$.
The SIT mode takes place when either $\Delta\to0$ or $\Delta\to\Delta_0$; the
SIST mode, when either $\Delta\to\Delta_1$ or $\Delta\to\Delta_2$.
If $-A_2\omega_0=A_1^2/4$, then, obviously, $\Delta_1=\Delta_2=\Delta_0/2$.
The parameter domain corresponding to SIST lies in the neighborhood of point
$\Delta=\Delta_0/2$ in this case.
If $-A_2\omega_0>A_1^2/4$, then the roots $\Delta_1$, $\Delta_2$ are complex
quantities, and the curve of strong excitation consists of a single branch.
Under this condition, the SIST mode may not exist.
As in the case of an isotropic medium and a positively birefringent medium,
the parameter $\tau_p$ can have arbitrary values for pulses that cause the
largest change of the population of the quantum levels.
Figure~3 shows the curves of strong excitation corresponding to several values
of parameters of the medium.

The NNT, PNT, and EOT modes must be less expressed in negatively birefringent
media because of weaker effective anisotropy.
This behavior resembles one of ''dense'' media, except that the effective
anisotropy decreases with increasing absolute value of detuning in the present
case.

Rationally decreasing pulses (\ref{11}), (\ref{11a}) exist in negatively
birefringent media if $-A_2\omega_0<A_1^2$.
Detuning of these pulses can be positive only
($\tilde\Delta_1<\Delta<\tilde\Delta_2$).

\section{MEDIA WITH DIPOLE-DIPOLE INTERACTION}

In resent years, there was a growing interest of investigating the coherent
phenomena in the optically dense media$\mbox{}^{13-21}$.
The difference between the macroscopic field and microscopic one acting on a
separate particle has to be taken into account at the description of the
interaction of radiation with the quantum systems in such the media.
The effect of the local field leads, for example, to the internal optical
bistability$\mbox{}^{15}$ and to the upconversion of the emission in dimeric
structures$\mbox{}^{16}$.
A consideration of near electric dipole-dipole interaction of the particles 
causes the dynamic shift of the frequency of resonant transition, which 
depends on the density population of levels$\mbox{}^{14}$.
In that case, the propagation of the electromagnetic pulses in optically dense
isotropic two-level media is described in the frameworks of the slowly varying
envelope approximation by the equations (\ref{1}), (\ref {3}) and by the
equation
\begin{equation}
\frac{\partial R}{\partial t}=i\left(\Delta+B(W-W_{\infty})\right)R+
i\Omega_oW.
\label{12}
\end{equation}
Here $B$ is the phenomenological parameter taking account of the interaction
between dipoles; variable $\Omega_o$ has a meaning of the Rabi frequency of
the complex envelope of the electric field of the pulse.

The results obtained in$\mbox{}^8$ and in the previous sections are extended
on the case of the optically dense media as follows.
Let us assume $n_e=n_o$.
Then the pulse solutions of the equations (\ref{1}), (\ref{3}), (\ref{12}) are
given by formulas (\ref{5}--\ref{9}) and (\ref{11}), (\ref{11a}) if we put
$$
\tilde D=2BW_{\infty}\frac{\omega_0\tau_p^2d^2}{1+\alpha^2},
$$
and
$$
\tilde D=2BW_{\infty}\frac{\omega_0d^2}{\Delta^2}.
$$
respectively.
Hence, the electromagnetic pulses can propagate through the optically dense
media in the transparency modes classified in$\mbox{}^8$.
The solution (\ref{11}), (\ref{11a}) describing the propagation of rationally
decreasing pulse of the self-induced transparency seems to be new.
It exists when detuning of the pulse satisfy condition $BW_{\infty}/\Delta>0$.
The rational form of the decay of the free polarization for such the media was
revealed in$\mbox{}^{21}$.

\begin{figure}[h]
\centering
\vskip-1.0cm
\includegraphics[angle=-90,width=4.0in]{A_00.eps}
\begin{picture}(0,0)
\put(-170,-44){\small I}
\put(-194,-180){\small II}
\put(-135,-145){\small III}
\put(-266,-176){\small V}
\put(-49,-176){\small IV}
\put(-272.5,-187.5){\vector(1,0){240}}
\put(-144.1,-187.5){\vector(0,1){154}}
\put(-38,-200){$\Delta$}
\put(-146.5,-200){0}
\put(-138,-42){$\tau_p^{-1}$}
\put(-180,-187.5){\line(0,1){150}}
\put(-180.3,-187.5){\line(0,1){150}}
\put(-179.7,-187.5){\line(0,1){150}}
\put(-198,-200){$BW_{\infty}/2$}
\end{picture}
\vskip-0.5cm
\centerline{
\parbox{14.0cm}{\small\bf Fig. 4. \rm Curve of the strong excitation for
optically dense medium. Parameter domains are shown for $B>0$ and labelled as
on Fig.~1.}}
\vskip-0.2cm
\end{figure}

It is interesting, that the largest excitation of the medium will be caused in
accordance with formula (\ref{10}) by the pulses having constant detuning:
$\Delta=BW_{\infty}/2$ (see Fig.~4).
The pulses can propagate in this case in the SIT mode ($\tau_p\to0$) or in
SIST one ($\tau_p\to\infty$).
If parameter
$$
g=\frac{1+\alpha^2}{BW_{\infty}\tau_p}
$$
satisfy condition $|g|\ll1$ (extremely dense media), then pulses can propagate
in the EOT mode ($|\alpha|\ll1$), NNT or PNT modes ($|\alpha|\gg1$).
The domains corresponding to different modes of the optical transparency are
given on Fig.~4.

The term responsible for the dipole-dipole interaction can be entered into
equations (\ref{1}--\ref{4}) also.
The solutions of the corresponding system follow from ones presented in
Section 2 after suitable redefinition of coefficient $\tilde D$.

\section{CONCLUSION}

The analysis of the solutions of system of material and wave equations 
(\ref{1}--\ref{4}) shows that the modes of the pulse propagation through the 
anisotropic media described in$\mbox{}^{7,8}$ exist in more general physical 
situation, when the approximation of the unidirectional propagation is 
inapplicable, and linear velocities of ordinary and extraordinary components 
are different.
The anisotropy of the medium becomes effective now, i.e. its degree depends on 
the parameters of the pulse also. 
Another important feature of the anisotropic media is that the rationally
decreasing pulses of the self-induced transparency can propagate through it.

If the concentration of the particles is not small, or the medium is 
negatively birefringent, then the effective anisotropy decreases.
For this reason, the modes of the transparency, which are accompanied by the
strong excitation of the resonant particles, become expressed.
As in the case studied previously$\mbox{}^8$, the carrier frequency of the 
ordinary component of the pulses in these modes is less than the frequency of 
resonant transition.
Under this condition only, the extraordinary component of the pulses can shift 
the energy levels and, simultaneously, generate the phase modulation in such a 
manner, that the ordinary component turns out to be in the resonance with the 
medium in an average and causes its strong excitation.

The more interesting case takes place when the concentration of the particles
is small, and the medium possesses the positive birefringence.
Its distinctive feature is that the effective anisotropy can change the sign.
As a result, the modes of the strong excitation exist not only if the carrier 
frequency of ordinary component is less than resonant one, but if it exceeds 
the resonant frequency also.
Besides, the degree of the effective anisotropy becomes unbounded, when the
pulse group velocity tends to linear velocity of the extraordinary wave.
Such the pulses propagate in the mode of the extraordinary
transparency$\mbox{}^7$.

The transparency modes introduced for the anisotropic media exist as well in 
isotropic, optically dense media, in which the direct electric dipole-dipole 
interaction between the resonant particles is considered$\mbox{}^{14}$.
The role of the dipole-dipole interaction coincides here with one of the 
extraordinary component in anisotropic case.

\section{ACKNOWLEDGMENT}

This work was supported by the RBRF grant 05--02--16422\,a.

\vfill
\eject
\end{document}